\documentclass[aps,reprint,10pt,twocolumn,superscriptaddress]{revtex4}
\usepackage{amsmath}
\usepackage{amssymb}
\usepackage{bbm}
\usepackage{graphicx}
\usepackage[countmax]{subfloat}
\usepackage{framed}
\usepackage{color}
\usepackage{hyperref}
\usepackage{epstopdf}
\usepackage{dsfont}
\usepackage{amsthm}
\usepackage{wrapfig}
\usepackage{relsize}
\usepackage{bm}
\usepackage{enumitem}
\usepackage{soul}

\newcommand{\red}[1]{{\color{red}}}

\usepackage{pifont}

\newcommand{\ba}{\begin{eqnarray}}
\newcommand{\ea}{\end{eqnarray}}

\begin{document}

\title{Adaptive Tracking of Enzymatic Reactions with Quantum Light} 

\author{Valeria Cimini}
\affiliation{Dipartimento di Scienze, Universit\`{a} degli Studi Roma Tre, Via della Vasca Navale 84, 00146, Rome, Italy}
\author{Marta Mellini}
\affiliation{Dipartimento di Scienze, Universit\`{a} degli Studi Roma Tre, Via della Vasca Navale 84, 00146, Rome, Italy}
\author{Giordano Rampioni}
\affiliation{Dipartimento di Scienze, Universit\`{a} degli Studi Roma Tre, Via della Vasca Navale 84, 00146, Rome, Italy}
\author{Marco Sbroscia}
\affiliation{Dipartimento di Scienze, Universit\`{a} degli Studi Roma Tre, Via della Vasca Navale 84, 00146, Rome, Italy}
\author{Livia Leoni}
\affiliation{Dipartimento di Scienze, Universit\`{a} degli Studi Roma Tre, Via della Vasca Navale 84, 00146, Rome, Italy}
\author{Marco Barbieri}
\affiliation{Dipartimento di Scienze, Universit\`{a} degli Studi Roma Tre, Via della Vasca Navale 84, 00146, Rome, Italy}
\affiliation{Istituto Nazionale di Ottica - CNR, Largo Enrico Fermi 6, 50125, Florence, Italy}
\author{Ilaria Gianani}\email{ilaria.gianani@uniroma3.it}
\affiliation{Dipartimento di Scienze, Universit\`{a} degli Studi Roma Tre, Via della Vasca Navale 84, 00146, Rome, Italy}
\affiliation{Dipartimento di Fisica, Sapienza Universit\`{a} di Roma, P.le Aldo Moro, 5, 00185, Rome, Italy}

\begin{abstract} 
Enzymes are essential to maintain organisms alive. Some of the reactions they catalyze are associated with a change in reagents chirality, hence their activity can be tracked by using optical means. However, illumination affects enzyme activity: the challenge is to operate at low-intensity regime avoiding loss in sensitivity. Here we apply quantum phase estimation to real-time measurement of invertase enzymatic activity. Control of the probe at the quantum level demonstrates the potential for reducing invasiveness with optimized sensitivity at once. This preliminary effort, bringing together methods of quantum physics and biology, constitutes an important step towards full development of quantum sensors for biological systems.
\end{abstract}

\maketitle
Enzymes are proteins acting as bio-catalyzers. They make  the vast majority of the chemical reactions required to sustain life possible~\cite{Cooper2000}. Beyond their fundamental importance in the physiology and metabolism of all living organisms, enzymes can be purified and engineered and they are largely used in industrial applications for food, chemicals, drugs and energy production. The stereo-selective production of chiral compounds is among the most studied and appreciated enzymatic activities, due to their use in the production of fine-chemicals in cost effective, nontoxic and eco-friendly processes~\cite{Choi2015}. 
Reaction kinetics of enzymes is usually monitored by measuring in time the conversion of the substrate molecules into the product molecules. This could be easily done by optical methods when studying enzymatic reactions in which the substrate and/or the product absorb light at a typical wavelength. In some applications, chromogenic reagents reacting with the substrate and/or the product can also be used. However, these procedures require to collect assays at different times during the reaction. The coarseness of the measurement however does not allow real-time monitoring~\cite{Cooper2000, Harris2009}.

Reactions leading to a change in optical activity of the enzymatic substrate with respect to the products can be monitored by circular dichroism or optical activity measurements with good accuracy, precision, and time resolution at moderate illumination, allowing real-time tracking of products accumulation~\cite{Harris2009, Oppermann2019}. Right- and left-circularly polarized light propagates with different refractive indexes, which cause one polarization to accumulate a small delay with respect to the other. This is equivalent to light traversing two different paths, with a relative optical phase due to birefringence. Real-time polarization analysis allows retrieving the value of such phase, hence enzymatic activity can be probed by measuring phase variation in time.
 
\begin{figure}
\centering
\includegraphics[width=.9\columnwidth]{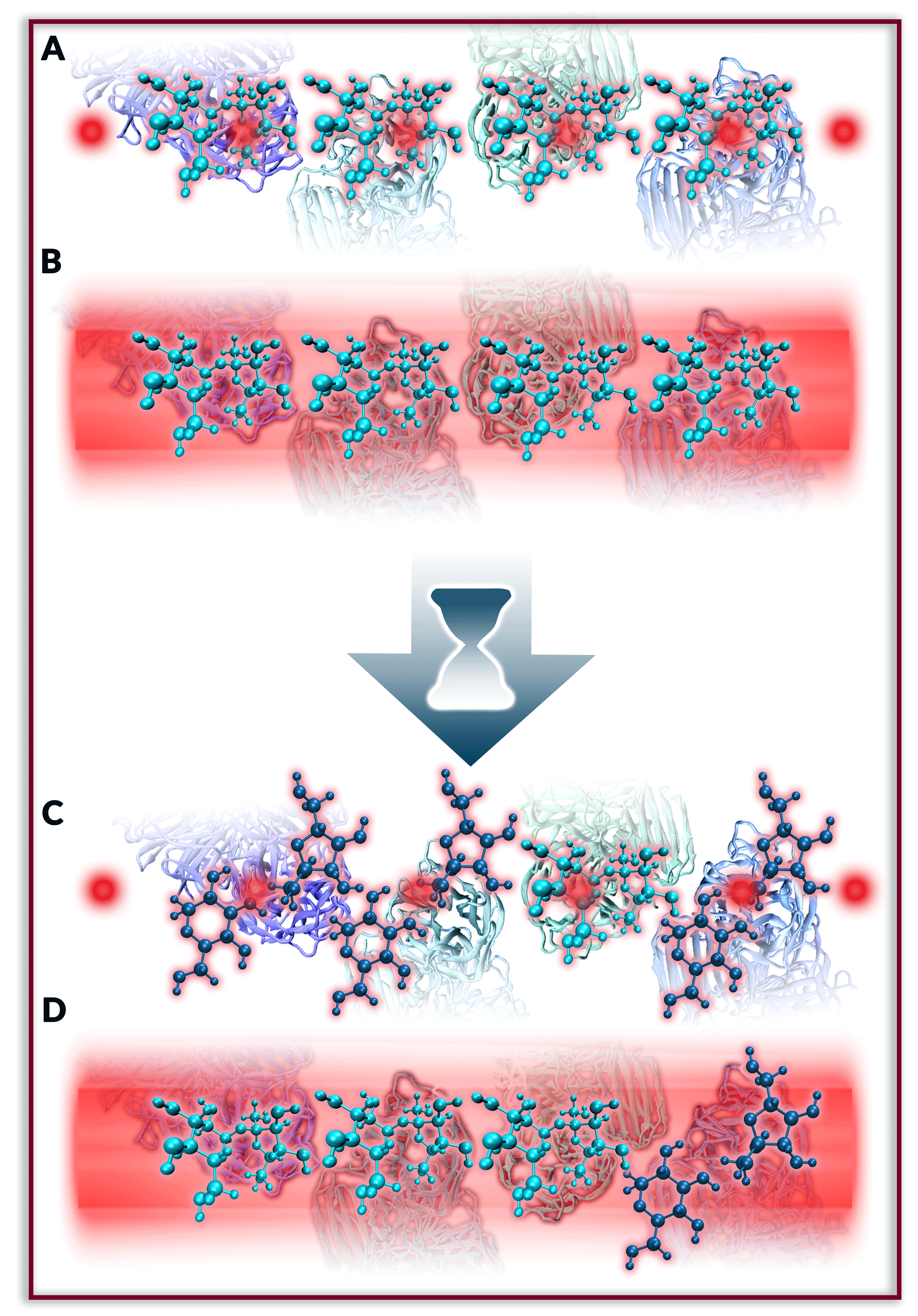}
\caption{Schematic rendition of the inhibitory effect of light on invertase (PDB: 4EQV) activity. Sucrose (light blue) is hydrolyzed to {\sc d}-glucose and {\sc d}-fructose (dark blue), while an optical measurement is carried out in either the low photon number (A) or intense laser regime (B). As the reaction is tracked, low illumination levels have a reduced impact on the enzymatic activity (C), while intense light could  altered it, due to thermal and/or electrical processes (D). However, limiting the intensity of light to which the sample is exposed would result in loss of precision. This can be recovered by engineering the quantum state of the probing light.}
\label{fig:figure1}
\end{figure}

The quality of an optical measurement is embodied as the uncertainty on the target parameter. This is influenced by the amount of light used for monitoring: increasing intensities is the primary way to decrease uncertainty, especially when the collection time is limited. However, working under high intensity regime may come at a dear price when exploring biological samples. Indeed thermal and electrical processes associated with exposure to intense light can affect protein properties (Fig.~\ref{fig:figure1}), or even cause their permanent damage~\cite{Carlton2010, Pena2012, Vojisavljevic2007, Mirmiranpour2018}. Improving precision could be worthless if the response is driven far from that of the unperturbed system. This originates a necessary trade-off between the modification of the enzyme activity and the quality of the measurement. Low-invasiveness optical probing would then be a practical solution in terms of ease of operation while maintaining time-tracking capabilities. However, the impact of shot noise becomes more relevant in the low light regime, compromising the signal-to-noise ratio.
 
This limitation can be overcome by engineering fundamental features of light at the quantum level. Quantum metrology is the art of identifying how such quantum properties need being controlled and measured, and provides clear guidelines on preparing the best possible probe for a given intensity~\cite{Dowling2008, Giovannetti2011, Giovannetti2006}. Superior accuracy is possible thanks to the careful control of the average intensity and of quantum fluctuations.
 
Here we apply quantum metrology for monitoring the activity of a bio-catalyzer, providing indications on an alteration of its activity due to laser light illumination; our proof of principle experiment supports the appropriateness of the low-invasiveness quantum metrology approach. We perform our study on invertase, a model enzyme for biochemistry since the birth of this discipline. Carbohydrates, including the di-saccharide sucrose, are a primary source of energy in living organisms, hence the sucrose hydrolyzing enzyme invertase plays a central role in cellular metabolism. The name invertase came from the inversion of the optical activity of a sucrose solution upon the enzymatic hydrolysis into the mono-saccharides {\sc d}-glucose and {\sc d}--fructose~\cite{Cooper2000}.

\section*{Results}
Monitoring biological samples with minimal invasiveness has been one of the main motivations behind the emergence of quantum metrology~\cite{Dowling2008, Giovannetti2011, Giovannetti2006}. Controlling light at the single-photon level offers the possibility of maintaining good signal-to-noise ratios, even in the low illumination regime. When probing optical activity, a fundamental relation exists, analogous to Heisenberg's uncertainty relation, linking the best possible uncertainty $\Delta \phi$ on the optical phase $\phi$ and the mean-square difference $\Delta N$ between the number of photons, i.e. light energy quanta, in the two polarisations~\cite{Giovannetti2006}: $\Delta\phi\, \Delta N\geq1/\sqrt{M}$  where $M$ is the number of repetitions of the experiment. The higher the difference  $\Delta N$ , the lower the achievable uncertainty  $\Delta \phi$ .

With classical sources, there is no control on the number of photons other than their average value, corresponding to the classical intensity. In the absence of any source of technical noise, when $N$ photons on average are distributed evenly between the two polarisations, the fluctuation is $\Delta N = \sqrt{N}$ , hence the phase uncertainty is $\Delta\phi=1/\sqrt{NM}$~\cite{Giovannetti2011, Giovannetti2006}. When the number of possible repetitions is limited, improving the precision on the phase demands higher intensity, i.e. average number of photons.

This condition is far from being optimal: instead,  is maximized by inputting exactly $N$ photons in the right-circular polarisation, and none on the other, and, {\it at the same time}, $N$ photons in the left-circular polarisation, and none on the other. In this way, probing occurs on both polarisations and achieves the optimal uncertainty $\Delta\phi=1/(N\sqrt{M})$, given the number of photons available~\cite{Dowling2008, Giovannetti2006}. This takes advantage from state of quantum superposition of all photons being in either polarisation at once: such states of light are called N00N states, and can provide a substantial advantage when simply augmenting the average photon flux is not a viable option.

N00N states have been reliably produced~\cite{Walther2004, Mitchell2004}, and have found several applications in phase measurements \cite{Nagata2007, ThomasPeter2011, Matthews2011, Xiang2011, Ono2013, DAmbrosio2013, Slussarenko2017, tipo} , including the study of light-sensitive samples~\cite{Wolfgramm2013}. These have been applied to optical activity dispersion of sugar solutions~\cite{Tischler2016}, and to the measurement of protein concentration in microfluidic channels~\cite{Crespi2012}, while a different mechanism for noise reduction (viz. amplitude squeezing) has allowed tracking the movements of lipid granules in cytoplasm with improved sensitivity~\cite{Taylor2013}.

We have applied N00N states with $N=2$ to monitoring the kinetics of invertase in time. These are generated by means of two-photon quantum interference. We start with a pair of photons, one in the horizontal (H) polarization and the other in the vertical (V) polarization, which are then combined on a polarizing beam splitter in order to make them indistinguishable in their spatial and temporal degrees of freedom (Fig. 2A). Indistinguishability ensures that in the circular polarizations the behavior is that of a N00N state (Fig. 2B). By passing through the sample, the photon pair accumulates an optical phase twice as fast as a single photon would, hence the relative phase between the two polarization states at the output is $2\phi$  (Fig 2C). It is possible to estimate the accumulated phase $\phi$, since this results in a rotation by an angle $\phi/2$ of the two original polarizations, which can be easily measured. The setup sketched in Fig. 2A allows collecting interference patterns as that in Fig. 2D. The visibility of such oscillations, is mostly limited by the spectral properties of the two photons.

\begin{figure}
\centering
\includegraphics[width=\columnwidth]{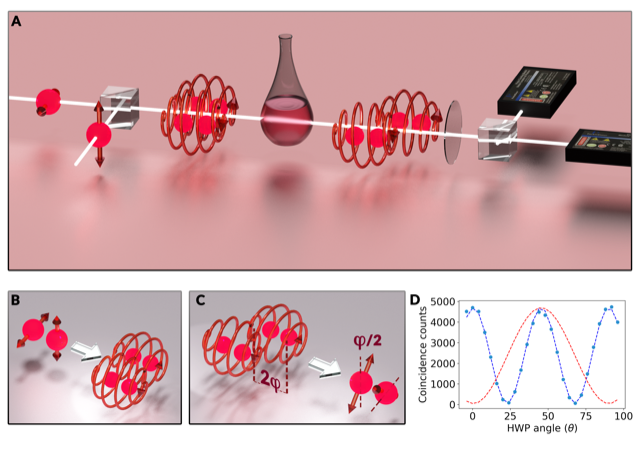}
\caption{Schematics of the experiment and its working principles. (A) Experimental apparatus used for the phase tracking with quantum light. Two photons produced via spontaneous parametric down conversion with orthogonal polarizations are sent on a polarizing beam splitter (PBS) in order to produce a N00N state - a quantum superposition in which both photons are in either left-circular or right-circular polarization at once (panel B). As the probe interacts with the sugar solution an optical phase $2\phi$  between the two components of the N00N state is introduced. When reverting to the linear polarizations, the phase shift corresponds to a rotation by an angle $\phi/2$  on both photons (panel C). Finally, polarization is analyzed by means of a half wave plate (HWP) and a second PBS, and the photons are detected with avalanche photodiodes, registering coincidence events. A calibration of the measurement apparatus is reported in panel D for no sample ($\phi=0$). Oscillations in the coincidences (blue dots) occur at twice the frequency than a standard probe would deliver (red curve). This grants an improvement in its metrological features. For further experimental details refer to the Supplementary Information.}
\label{fig:figure2}
\end{figure}

Differently from previous investigations, here a non-stationary optical phase, whose evolution reflects enzymatic activity, is addressed. Such time-tracking task poses distinctive requirements on its implementation with respect to the stationary case. As tracking takes place over a long interval, the measuring system might be prone to drifts and optical misalignments, which can be further aggravated by spurious effects in unstable samples, as biological specimens. Therefore, continuous monitoring of the quality of the probe state while performing the probing is needed, since the non-stationary conditions would limit the reliability of any pre-calibration. In the language of quantum metrology, this is cast as a problem of multiple parameter estimation, where one parameter is the optical phase, as in conventional estimation, and the other is a quality figure for the probe, in our case, the oscillation visibility. In our investigation, we adopt the methodology developed in Ref.~\cite{Roccia2018}, that has been validated by observing sucrose hydrolysis catalyzed by hydrochloric acid~\cite{Cimini2019}.

Critically, the achievable uncertainty $\Delta\phi$  depends on the actual value of $\phi$: as this varies over time, optimality conditions can be lost. If the evolution explores a limited range of phases, as in Ref.~\cite{Cimini2019}, the working conditions could still be favorable. In the general case, an adaptive procedure should be carried out in such a way that the measured quantity is a small displacement $\delta\phi=\phi-\phi_p$  from a predicted phase $\phi_p$, and the optimal measurement is performed for $\delta\phi=0$. This technique has been successfully applied to quantum squeezed light~\cite{Yonezawa2012, Berni2015}, and we extend it here to N00N states.

\begin{figure}
\centering
\includegraphics[width=\columnwidth]{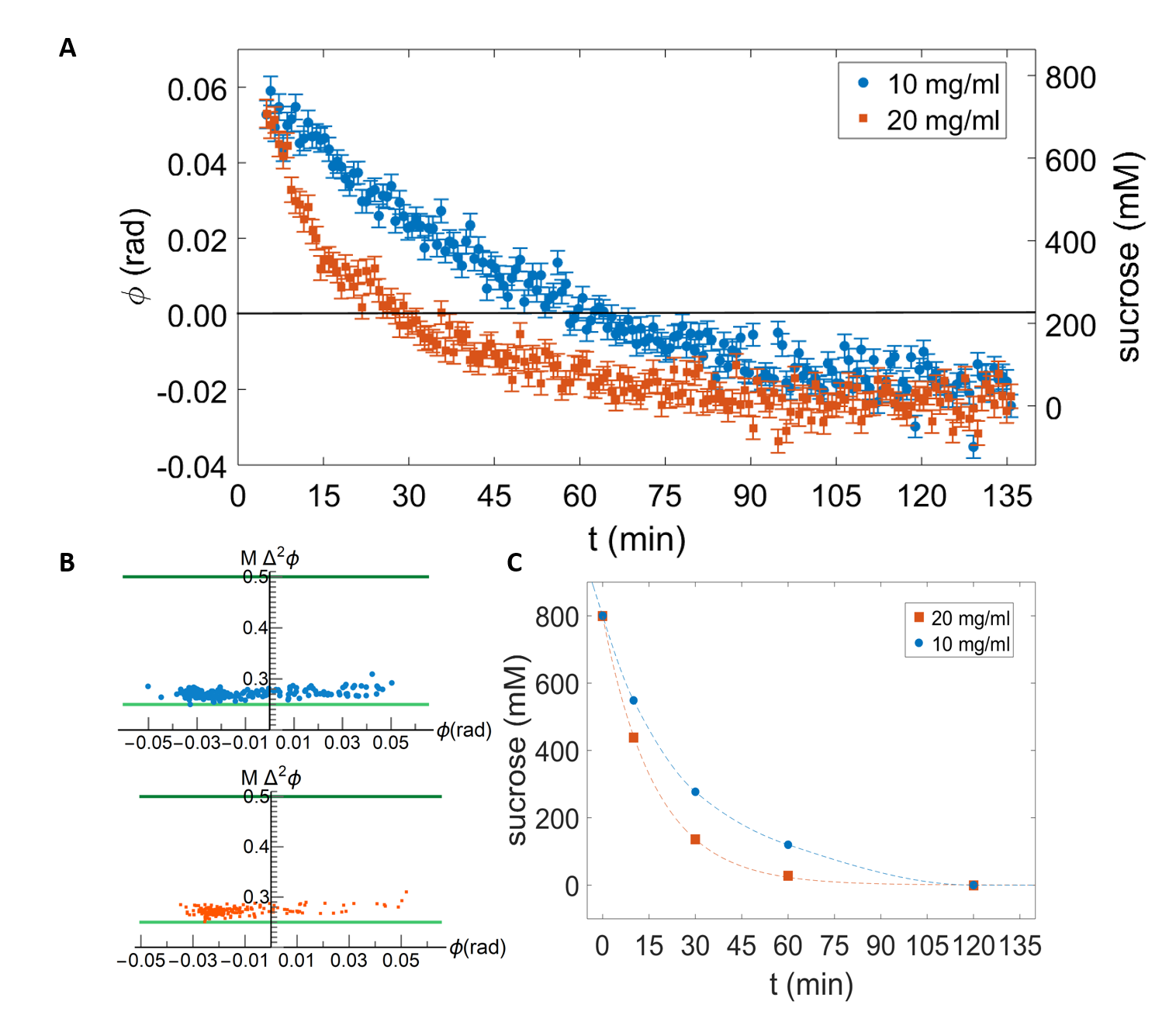}
\caption{Experimental results of optical tracking and DNS assays. (A) Tracking of invertase activity using N00N states, for two different enzyme concentrations: 10 mg/ml (blue dots) and 20 mg/ml (orange squares). The concentration of sucrose corresponding to each phase measurement is also shown on the right axis. The sampling interval is 37 s. (B) Errors obtained with the adaptive measurement strategy for the two concentrations (color scheme as panel A). The light green line corresponds to the optimal lower bound achievable with 2-photon N00N states, while the dark green line is the bound related to a classical probe with the same average intensity. (C) Tracking with the DNS assay for two different invertase concentrations: 10 mg/ml (blue dots) and 20 mg/ml (orange squares).}
\label{fig:figure3}
\end{figure}

The experiment proceeds as follows: first, a solution containing sucrose (0.8 M) is prepared; then invertase is added to the sample and this sets the initial time for the time-tracking measurement. We record the kinetics at room temperature with two different invertase concentrations, 10 mg/ml and 20 mg/ml. We estimate one value for the phase and one for the visibility with a sampling rate of 37 s. In Fig.~\ref{fig:figure3}A we show how the phase evolves in time due to the catalyzing action of invertase: the original positive phase of the sample is gradually modified to the negative value of the final products. The enzyme concentration dictates the time scale of the kinetics. For each point, the choice of the measurement is optimized with the aforementioned adaptive scheme to ensure that each phase is estimated with an uncertainty close to the ultimate limit allowed by the number of collected events and by the number of photons per event (Fig.~\ref{fig:figure3}B). These post-selected measurements do not take into account the impact of losses, which is discussed below.

The optical measurements have been validated by using a standard method based on dinitrosalicylic acid (DNS). When DNS binds to {\sc d}-glucose and {\sc d}-fructose, it shows a typical absorption peak at 540 nm wavelength, whose intensity is proportional to the amount of these mono-saccharides~\cite{Sumner1921, Combes1983}. The same time scales for the completion of the process are observed with the two techniques, with a variabilit that could be attributed to small variations in room temperature and/or invertase and sucrose batches concentration (Fig.~\ref{fig:figure3}C). 

To investigate possible effects of light on invertase activity, additional reactions were carried out with invertase samples illuminated for 1 h with lasers at different frequencies and intensities. 

\begin{figure}
\centering
\includegraphics[width=\linewidth]{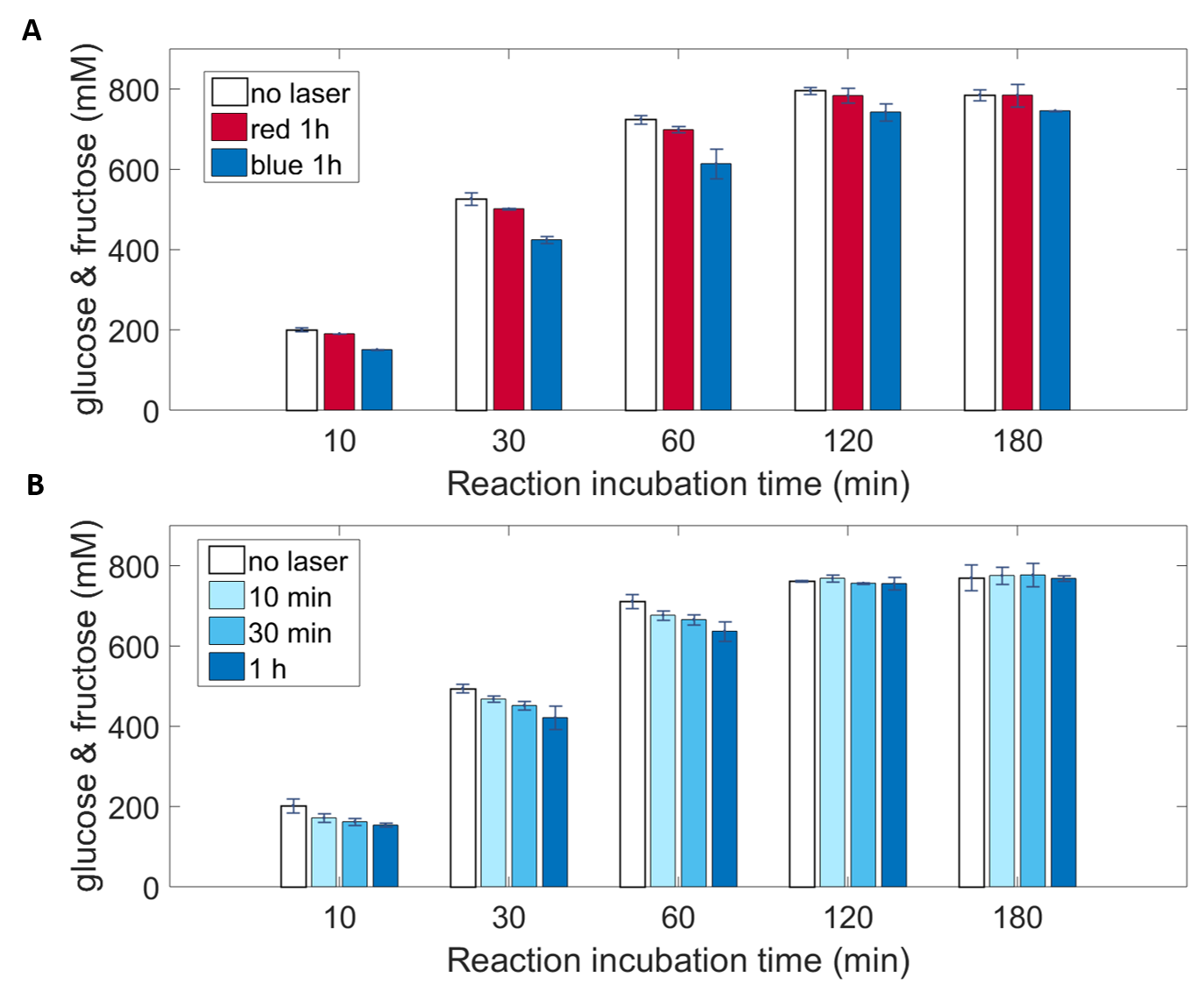}
\caption{Tracking invertase activity after exposure to laser light. (A) Products concentration measured with the DNS assay for samples of sucrose solution catalyzed by 10 mg/ml invertase undergone different illumination conditions. White: control sample; red: 1 h illumination with a 2.6 mW 800 nm CW laser; blue: 1 h illumination with 200 mW 405 nm CW laser. (B) Tracking with DNS assay performed on 0.8 M sucrose solutions mixed with 10 mg/ml invertase solutions undergone different illumination times. White: control sample (not illuminated); cyan: invertase illuminated for 10 min; light blue: invertase illuminated for 30 mins; dark blue: invertase illuminated for 60 min.}
\label{fig:figure4}
\end{figure}

Three 70 $\mu$l samples of 10 mg/ml invertase solution were incubated for 1 h with no laser illumination or illuminated with a 2.6 mW CW laser at 800 nm (25 mW/$cm^2$), or with a 200 mW CW laser at 405 nm (2 W/$cm^2$). These two itensities have been chosen to address common fluencies in optical measurements as well as a regime known to be disurptive. Then, 20 $\mu$l of these invertase solutions (no laser, red 1 h and blue 1 h, respectively) were added to 2 ml of 0.8 M sucrose solution to start invertase reactions ($t = 0$). Invertase activity in these samples was monitored at different time points by means of the DNS method. Comparison to untreated sample (i.e. not illuminated invertase) revealed that light exposure is detrimental to enzymatic activity (Fig.~\ref{fig:figure4}A). A reduction of the activity up to 5\% is observed after illumination at 800nm. This increases to 25\% with the 405nm laser. As expected, the difference with respect to the untreated sample is more pronounced in the early stages of the catalysis, since this originates in the altered time scale for the reaction. In order to inspect the laser-mediated alteration of invertase activity at different exposure times, four 70 $\mu$l samples of 10 mg/ml invertase solution were incubated for 1 h without laser illumination or illuminated with 200 mW CW laser at 405 nm for 10 min, 30 min, or 1 h (no laser, 10 min, 30 min and 1 h, respectively). Also in this case, invertase activity in the samples was monitored by means of the DNS method. The results are shown in Fig.~\ref{fig:figure4}B.

\section*{Discussion}
Optical activity is a precious and reliable tool for inspecting enzymes activity and, adopting quantum metrology for its investigation grants minimal disruption during tracking. Future perspectives include the combination of polarization-resolved measurements with imaging techniques to map the kinetics both in space and time.

Notably, this proof of principle study is not meant to provide an unconditional advantage with respect to the ideal classical case with the same average intensity. Our goal is to demonstrate how quantum metrology techniques can be applied to this important class of problems and this has been successfully accomplished. 

The main obstacle towards a genuine quantum advantage in measurements is  achieving sufficiently low losses either due to low-efficiency detectors or to absorption of the samples themselves. The presence of losses has to be taken into account as it strictly binds the improvement in the phase uncertaintiy attainable with quantum resources to strong constraints for the visibility of the setup, eventually preventing any advantage even with perfect visibility in extremely lossy scenarios  \cite{opticaren, timerev}.  An improvement in the uncertainty is achieved only if the visibility satisfies:

\begin{equation}
V>\sqrt{\frac{1}{\eta\,N}},
\end{equation}

where $\eta$ encompasses all losses regardless their physical origin. This sets an ultimate bound for the maximum amount of losses that can be suffered; in our experiment this amounts to $\eta \simeq 50\%$. 
We show that by considering only post-selected measurements, our visibility is such that we do attain an improvement compared to classical resources, however securing an improvement on non-post-selected data remains elusive since our overall efficiencies is $\eta \simeq 5\%$, which is also limited by the quantum yield of the detectors as well as pronounced absorption form water in the sample.  
In this respect single photon detectors are now able to operate in the appropriate regime of efficiency, these are optimized for functioning in the near infrared range~\cite{Slussarenko2017}. On the other hand, these wavelengths are less sensitive to optical activity effects in biological materials, due to the reduction of the refractive indexes.  Future perspectives should take up the challenge of developing full systems, from emission to detection, able to operate in the visible range, taking also into account the absorption features of water. 

Optimizing the optical setup, the biological samples preparation, and the validation procedure has dictated the multidisciplinary approach in our study. As biological applications of quantum metrology are among the most ambitious and rewarding, it is crucial to develop tools and methods through which the communities can establish a common ground. Our research is a first attempt in this direction.

\section*{Supplementary Information}

\subsection*{Reagents and samples preparation}
Invertase from {\it Saccharomyces cerevisiae} (Sigma-Aldrich Inc.), was dissolved at 10 mg/ml or 20 mg/ml concentrations in 0.1 M acetate buffer pH 4.5 [11.96 g/L sodium acetate (Fluka BioChemika), 2.22 g/L acetic acid (Carlo Erba reagents)]. Sucrose (Sigma-Aldrich Inc.), {\sc d}-glucose (Fluka BioChemika) and {\sc d}-fructose (Fluka BioChemika) were dissolved at 0.8 M concentration in 0.1 M acetate buffer pH 4.5. The dinitrosalicylic acid (DNS) reagent~\cite{Sumner1921} was prepared by dissolving 10 g of DNS (Sigma-Aldrich Inc.) and 300 g of sodium potassium tartrate (Riedel-de Ha\"en) in 800 ml of 0.5 N NaOH (Sigma-Aldrich Inc.), then the volume was made up to 1.0 L with distilled water.
Samples were prepared by adding 20 $\mu$l of invertase solution (10 mg/ml or 20 mg/ml concentrations) to 2 ml of 0.8 M sucrose solution. Twenty-$\mu$l of 0.1 M acetate buffer pH 4.5 were added to 2 ml of 0.8 M sucrose solution as blank sample (control sample without invertase).
These samples were used for tracking invertase activity both with quantum light and with the DNS method. All reactions were performed at room temperature.
 
\subsection*{Invertase activity assay (DNS method)}
Invertase-dependent accumulation of the sucrose hydrolysis products {\sc d}-glucose and {\sc d}-fructose was monitored at different time points via the DNS method~\cite{Combes1983}, with minor modifications. One hundred-$\mu$l of the reaction mixture were added to 200 $\mu$l of DNS reagent. Samples were boiled for 5 min and then cooled on ice for 5 min. One hundred-$\mu$l samples were dispensed into 96-wells microtiter plates, and absorbance at 540 nm ($A_{540}$) was measured with a Spark 10M luminometer-spectrophotometer (Tecan). A standard curve generated with serial dilutions of {\sc d}-glucose and {\sc d}-fructose (from 0.39 mM to 800 mM) was used to estimate the concentration of these monosaccharides in the reaction samples. For {\sc d}-glucose/{\sc d}-fructose concentrations $\geq$ 12.5 mM, samples were diluted in 0.1 M acetate buffer pH 4.5 to obtain $A_{540}$ values in the confidence detection range.

\subsection*{Optical setup}
Photon pairs are generated by spontaneous parametric down conversion (SPDC) source: a 3 mm $\beta$-Barium-Borate (BBO) nonlinear crystal is pumped with a CW laser at 100 mW at 405 nm (model Coherent OBIS). The laser beam is vertically polarized. The crystal is cut for a Type-I degenerate emission, so the generated photon-pairs are emitted at 810 nm and are  horizontally polarized. To ensure photon indistinguishability, the photons are selected through interference filters with a bandwidth of 7.5 nm and single mode fibers. 
The two photons are then prepared into a N00N state in the circular polarization by means of half wave plates (HWPs) and a polarizing beam splitter (PBS): the polarization of the signal photon remains unaltered (horizontal polarization) while a HWP at 45$^\circ$ rotates the idler photon polarization to vertical. Combining the two photons on a PBS results in the desired N00N state, with $N = 2$, in the circular polarization basis:
\begin{equation}
\begin{aligned}
|\psi\rangle&=\hat{a}^\dagger_H\hat{a}^\dagger_V|0\rangle\\
&= \frac{1}{2}\big((\hat{a}^\dagger_R)^2-(\hat{a}^\dagger_L)^2\big)|0\rangle\\
&= \frac{1}{\sqrt{2}}\big(|2_R0_L\rangle-|0_R2_L\rangle\big),
\end{aligned}
\end{equation}
where $H$ and $V$ indicate the horizontal and vertical polarization, while $R$ and $L$ the right- and left-circular polarization.
 
Passage through the chiral compound imparts a phase shift $2\phi$ on the $R$ polarization mode, leaving the other one unchanged. In the linear polarization basis this corresponds to imparting a rotation of   to each photon:
\begin{equation}
\begin{aligned}
|\tilde{\psi}\rangle&= \Big(\cos\Big(\frac{\phi}{2}\Big)\hat{a}^\dagger_H+\sin\Big(\frac{\phi}{2}\Big)\hat{a}^\dagger_V\Big) \times\\
&\Big(\cos\Big(\frac{\phi}{2}\Big)\hat{a}^\dagger_V-\sin\Big(\frac{\phi}{2}\Big)\hat{a}^\dagger_H\Big)|0\rangle\\
&= \cos\phi\big(\hat{a}^\dagger_H\hat{a}^\dagger_V|0\rangle\big)-\sin\phi\Big(\frac{(\hat{a}^\dagger_H)^2-(\hat{a}^\dagger_V)^2}{2}|0\rangle\Big).
\end{aligned}
\end{equation}

The sucrose solution prepared as described above, is placed in a Hellma Analytics quartz cuvette with an optical path of 2 cm. Adding invertase at the concentrations stated above sets $t = 0$ for the kinetic tracking. The sucrose and invertase solution were thoroughly mixed to avoid any refractive index gradients in the medium, which would cause major signal misalignment and loss. The procedure took up to 5 min, that is why the first measurement is recorded after that amount of time. This could be reduced further by means of automated stirrers.
 
The detection stage consists in a HWP and a second PBS, which are used to select different polarizations by rotating the HWP at an angle  . Photon counting is performed via single-mode-fiber-coupled avalanche photodiodes (APD) on each of the two output arms of the PBS. Coincidence counts between the two detectors are recorded. The electric signals converted by the APDs are then carried to the Field Programmable Gate Array (FPGA) board which delivers the coincidences counts, with the coincidence window fixed at 2 ns. Typical count rates are around 2000 coincidences/s, as reported in Fig. 2D.

\subsection*{Multiparameter estimation}
\begin{figure}
\centering
\includegraphics[width=\linewidth]{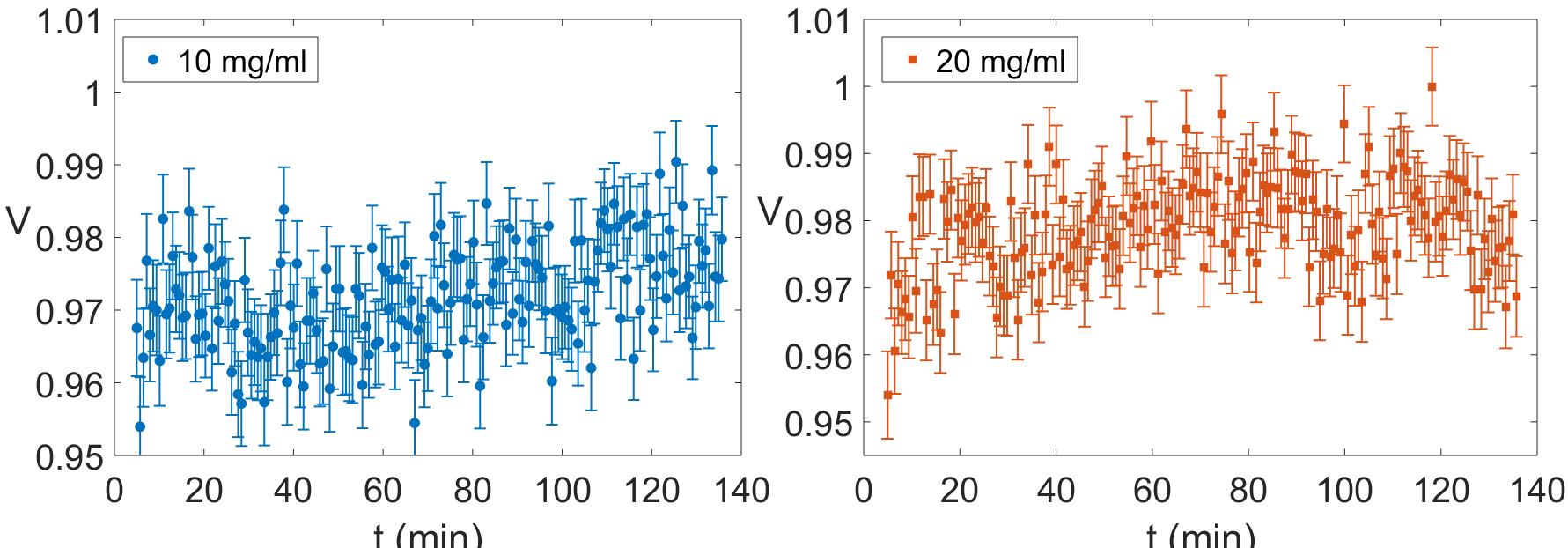}
\caption{Visibility measurements. The figures report the visibility values obtained for the 10 mg/ml (blue) and 20 mg/ml (orange) invertase multiparameter tracking. The measurements show that the visibility does vary in the time scale of the reaction process, hence demonstrating the need for the multiparameter approach to obtain an unbiased estimation of the optical activity in time.}
\label{fig:figure5}
\end{figure}

The correct estimation of the tracked phase is conditional on the correct evaluation of the setup and samples imperfections and instabilities that can vary with time. In particular, since we want to follow a dynamic process that can last hours, we need to address the fact that the level of noise itself can undergo considerable changes over such a long time span. The strategy we had adopted is to use the resources both for the phase estimation and for describing the quality of our probe state that investigates the sample simultaneously during the whole measurement process. In our case, the figure of merit  is the visibility $V$ of the two-photon interference given by the setup: the two-photon interference on the PBS occurs with a certain modulation depth {\it i.e.} the visibility, depending on how much the pair is indistinguishable. The interference fringes will then be a function of both $\phi$ and $V$ and an incorrect estimation of the latter would result in a bias on the estimation of the phase. In a situation that evolves in time it is of paramount importance to follow also the evolution of the visibility, and following~\cite{Roccia2018}, we have done so by adopting a multiparameter approach. 
For each data point we thus had obtained a value for the phase and a value for the visibility of the interferometer used for the phase evaluation. The values of the visibility corresponding to the phase dynamics during the reaction of sucrose with both the employed concentrations of invertase are presented in Fig.~\ref{fig:figure5}. The range of all the possible experimental imperfections which result in a reduced effective visibility is extremely differentiated, but we note that the instabilities of the sample itself are those that most affect their time dependence on the time-scales dictated by the reaction completion.
Taking into account the visibility parameter as well as the phase, the probabilities of detecting a coincidence event, selecting one of the angles $\theta\in\{\theta_0,\theta_0+\pi/16,\theta_0+\pi/8,\theta_0+3\pi/16\}$ by means of the HWP in the detection stage, are in the form:
\begin{equation}
p(\theta;\phi,V)=\frac{1}{4}(1+V\cos(8\theta+2\phi)).
\end{equation}
These probabilities are those post-selected on coincidences between the two detectors to account for the lack of photon-number-resolved detectors~\cite{Roccia2018, Cimini2019}. The value of $\phi$ and that of $V$ are obtained performing a Bayesian estimation starting from the four probabilities. 
The sucrose concentration $C_s(t)$ for a given phase measurement $\phi(t)$ can be easily calculated with the following~\cite{Cimini2019}: 
\begin{equation}
C_s(t) = C_s(t_0)\frac{\phi(t)-\phi(t_\infty)}{\phi(t_0)-\phi(t_\infty)},
\end{equation}
 where $\phi(t_0)$ and $ C_s(t_0)$  are the initial phase and concentration of sucrose, and $\phi(t_\infty)$ is that measured when the reaction has reached its completion.

\subsection*{Adaptive measurement calibration}
\begin{figure}
\centering
\includegraphics[width=\linewidth]{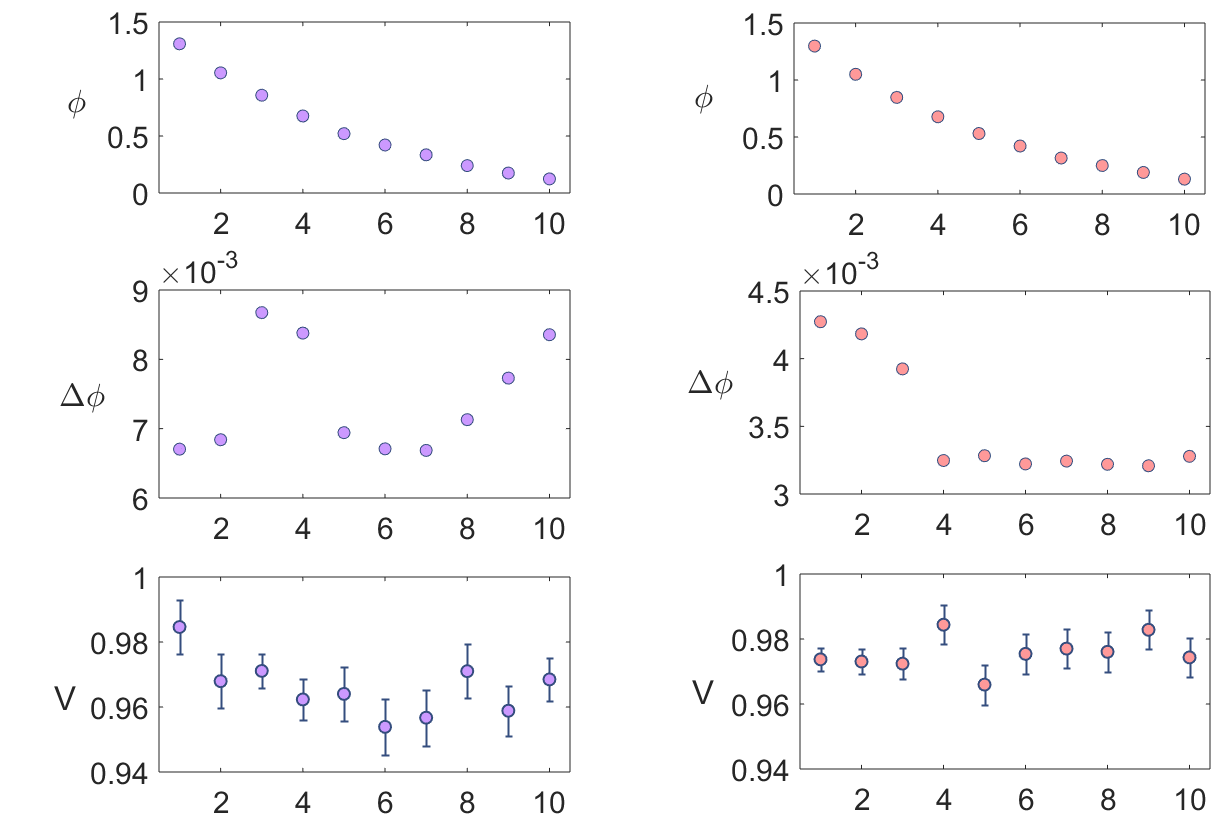}
\caption{Test of the adaptive measurement. The left column shows the measurement of a known phase obtained with a HWP inserted in place of the sample (upper panels) performed using the settings with $\theta_0=\pi/32$, while the right column shows the same measurement adapting the setting at each phase value; the prediction is obtained by a fitting of the previous values. After three points the error   drops to its minimum (center panels), which is then tracked through all the remaining phases. The visibility is also shown (lower panels).}
\label{fig:figure6}
\end{figure}

The uncertainty on the estimation of a phase $\phi$ is inversely proportional to the Fisher Information extracted by the measurement performed~\cite{Giovannetti2011, Giovannetti2006}. The Fisher Information is a matrix depending on the phase to be measured, on the parameter $V$, and on the measurement settings~\cite{Roccia2018}; this implies that the uncertainties on the phase will be phase-dependent as well. In order to minimize the uncertainty, a possible strategy is to perform an adaptive measurement so that the Fisher Information on the phase is always maximal. Optimal estimation for the setting $\theta_0=0$ is hence obtained for $\phi=\pi/8$.
Optimal estimation at the generic value $\phi$ demands to fulfill the condition $8\theta_0+2\phi=\pi/4$. This indicates that information on the value of the phase to be measured is required to accommodate the measurement accordingly. Hence, being able to predict the phase during the tracking is key to achieve near-optimal accuracy for each phase by choosing the appropriate value of $\theta_0$ .
The predictions can be obtained by either self-regression of the time sequence of the values of $\phi(t)$, possibly complemented by filtering~\cite{Yonezawa2012}: since the expected evolution of the phase is  $\phi_p(t)=\phi(t_0)e^{-(t-t_0)/\tau}$, this functional form was used to obtain a fitting curve, which was updated at each point. 
To test this approach, we performed a measurement inserting in the setup a calibrated known phase by means of another HWP. As shown in Fig.~\ref{fig:figure6}, this strategy allows keeping the uncertainty at its near-optimal level from the fourth phase value on. Naturally, the uncertainty on the first measurement is not optimized, since there is no prior information on the phase; the following two points are obtained via interpolation, and from the fourth point onwards the exponential fit gives a reliable estimation and can hence be used to accurately predict the successive phase value.

\begin{acknowledgments}
The authors would like to thank E. Roccia, J.P. Wolf and N. Treps. for fruitful discussion, and M.A. Ricci for lending chemical equipment. The Grant of Excellence Departments, MIUR-Italy (ARTICOLO 1, COMMI 314 - 337 LEGGE 232/2016) is gratefully acknowledged.
\end{acknowledgments}

\bibliography{cimini.bib}
\bibliographystyle{apsrev4-1}

\end{document}